\documentclass[twocolumn]{aa}

\usepackage{graphicx}
\usepackage{natbib}
\bibpunct{(}{)}{;}{a}{}{,}
\usepackage{txfonts}
\usepackage{xspace}

\graphicspath{{./}{figures/}}

\begin{document}

\title{The double white dwarf merger model for the progenitors of SN~2021yfj-like events}

   \author{Dongdong Liu\inst{1},
          Chengyuan Wu\inst{1},
          Takashi J. Moriya\inst{2,3,4},
          Jiawen Ning\inst{1},
          Zhengwei Liu\inst{1},
          Zhanwen Han\inst{1}
          \and
          Bo Wang\inst{1}
          }
   \institute{International Centre of Supernovae (ICESUN), Yunnan Key Laboratory of Supernova Research, Yunnan Observatories, Chinese Academy of Sciences (CAS), Kunming 650216, China\\
              \email{liudongdong@ynao.ac.cn; wangbo@ynao.ac.cn}
          \and
          National Astronomical Observatory of Japan, National Institutes of Natural Sciences, 2-21-1 Osawa, Mitaka, Tokyo 181-8588, Japan
          \and
          Graduate Institute for Advanced Studies, SOKENDAI, 2-21-1 Osawa, Mitaka, Tokyo 181-8588, Japan
          \and
          School of Physics and Astronomy, Monash University, Clayton, VIC 3800, Australia}

   \date{}

  \abstract
{Recently, a new class of supernovae (SN) with prominent narrow emission features of silicon (Si), sulfur (S) and Argon (Ar) has been reported, i.e. SN~2021yfj-like events (so-called ``SNe Ien''). Their progenitor origin is still unknown.
It has been suggested that a massive CO WD may evolve into a Si/S-rich WD when accreting He-rich matter at a high mass-transfer rate.
If the He companion subsequently evolves into another more massive WD, the merger of this double WD system can generate Si/S-rich circumstellar material through tidal stripping. Following the SN explosion, the interaction between the SN ejecta and the Si/S-rich circumstellar material could produce the Si, S, and Ar emission lines characteristic of SN~2021yfj-like events.}
{In this work, we aim to determine the initial parameter space of WD+He star systems that can lead to SN~2021yfj-like events via the double WD merger scenario, and to quantify their theoretical Galactic birthrate and delay-time distributions.}
{We perform detailed binary evolution simulations of a large number of semidetached WD+He star systems to obtain the parameter space that leads to the formation of Si/S-rich WDs and subsequent double WD mergers.
We then use binary population synthesis to calculate the Galactic birthrate and delay-time distribution of such events.}
{For the formation of SN~2021yfj-like events, we found that the initial CO WD and He companion masses must lie in the ranges of $1.0$--$1.2\,M_{\odot}$ and $2.2$--$2.5\,M_{\odot}$, respectively.
The derived merger rate for this scenario is $\sim(2.6\text{-}4.4) \times 10^{-5}\,\text{yr}^{-1}$ with delay times longer than 450 Myr, which is about 1\% of the observed SN Ia rate. }
{We suggest that the double WD merger scenario involving a Si/S-rich WD originating from a CO WD+He star system, represents a significant and competitive model for SN~2021yfj-like events, underscoring the need for further observations of similar events coupled with relevant theoretical investigations.}


\titlerunning{The double WD merger model for the progenitors of SN~2021yfj-like events}

\authorrunning{D. Liu et al.}

\maketitle
\nolinenumbers
 
\section{Introduction} 
Supernovae (SNe), the catastrophic explosions marking the terminal stellar evolutionary stages, are among the most energetic events in the universe (see \citealp{1960ApJ...132..565H,Parrent2014}).
Their study holds great significance across the stellar physics, galactic chemical evolution and cosmology studies (e.g. \citealp{Greggio1983,Matteucci1986,Riess1998AJ,Perlmutter1999ApJ,Cappellaro1999,Mannucci2006MNRAS,Maoz2014}).
SNe are broadly classified through a combination of observational characteristics and theoretical progenitor models. 
Observationally, the primary taxonomy is based on their optical spectra near peak brightness (\citealp{filippenko1997}).
According to the presence or absence of hydrogen (H) absorption lines, SNe can be distinguished into type II and type I, respectively.
Type I supernovae are further subdivided:
Type Ia SNe show strong silicon (Si) II absorption feature near their maximum luminosities without He lines, while Type Ib and Ic SNe show no Si absorption feature with significant helium (He) or Carbon-Oxygen (C-O), respectively.
In addition to this canonical classification, a growing number of supernovae exhibit narrow emission lines arising from the interaction of SN ejecta with pre-existing circumstellar material (CSM), as seen in Type IIn (H-rich; \citealp{Schlegel1990}), Type Ibn (He-rich; e.g., \citealp{Anupama2009}), and Type Icn (C/O-rich; e.g., \citealp{GalYam2022, Perley2022}) events (see also reviews by \citealp{Smith2017, Fraser2020}).
Theoretically, supernovae have been suggested to arise from two distinct physical mechanisms, i.e. the thermonuclear explosion of white dwarfs (WDs) in close binaries (corresponding to Type Ia SNe; for recent reviews see \citealt{Maoz2014,Wang2018RAA,Liu2023RAA....23h2001L,Ruiter2025A&ARv..33....1R}) and the core-collapse SNe resulting from the gravitational collapse of massive stars ($>8\,\rm M_{\odot}$; corresponding to Type II, Ib, and Ic SNe; for recent reviews see \citealt{Smartt2009ARA&A, Burrows2024ApJ...964L..16B}).

\citet{schulze2025} recently reported the discovery of SN~2021yfj that displays strong narrow Si, sulfur (S) and also argon (Ar) emission features, which reveals the interaction between the SN ejecta and a dense Si/S-rich CSM.
Following its spectral characteristics, \citet{schulze2025} proposed the designation “Ien” for this SN type, although its classification remains under discussion \citep{Moriya2026PASJ...78L..15M}.
Theoretically, it remains challenging to explain the origin of such a dense, Si/S-rich CSM. Nevertheless, the weak but detectable He emission lines also offer important constraints that can help distinguish between competing progenitor models (\citealp{schulze2025}).

The progenitor origin of the peculiar supernova SN~2021yfj remains under debate. \citet{schulze2025} proposed three possible models: (1) a pulsational pair‑instability SN from a high-mass massive star stripped down to the Si/S-rich layers \citep[e.g.][]{Woosley2007}, which can match the most features in the spectrum and light curve well, but struggles to account for the presence of He emission features; (2) a low‑mass massive star that sheds its envelope via silicon‑flash‑driven winds or a jet‑driven explosion \citep[e.g.][]{Woosley2015}, for which the physical details are highly uncertain and observationally poorly constrained; and (3) a compact-object merger (e.g., NS/WD with Si and S on its surface) that could supply Si/S-rich ejecta (\citealp{Khokhlov1986, Waldman2011}), but the production of Si and S are subdominant to the more common elements like Ca, Mg and Fe-group elements.

Motivated by this open question, we will focus on an alternative compact‑object merger channel—the double WD merger scenario (see \citealp{Moriya2026PASJ...78L..15M}).
In contrast to the surface He detonation on a low-mass He-accreting WD discussed by \citet{schulze2025} (see \citealp{Khokhlov1986, Waldman2011}), we consider the scenario where a massive CO WD accretes He-rich matter at a relatively high mass-transfer rate (e.g. \citealp{Ruiter2013MNRAS.429.1425R,Liu2016,Liu2018DD,Ruiter2019MNRAS.484..698R,Liu2020}). 
In this scenario, a sufficiently massive primary WD accreting at a sufficiently high mass-transfer rate will evolve into a Si/S-rich WD (maybe a hybrid WD with a CO core and a Si/S-rich shell or even a WD whose entire composition is Si/S-rich; e.g., \citealp{Nomoto1985ApJ...297..531N,Saio1985A&A...150L..21S,Saio1998ApJ...500..388S,Brooks2016ApJ...821...28B,Schwab2016MNRAS.463.3461S,wang2017,wu2019,wu2020}).
If the He star evolves into a more massive ONe WD, the gravitational wave radiation will drive the merger of this double WD system (a Si/S-rich WD and an ONe WD). The tidal disruption of the Si/S-rich WD both releases material to form a Si/S-rich CSM and leaves a Si/S-rich layer around the ONe WD (see \citealp{Moriya2026PASJ...78L..15M}).
The explosion of this configuration could potentially be a promising candidate for producing a SN with spectral characteristics similar to those of SN~2021yfj—namely, prominent strong Si, S and Ar emission lines, and potentially residual He features.
However, it remains unknown what initial parameter space of WD+He star systems can lead to such double WD mergers, and what the corresponding Galactic birthrate is.

In the present work, we will provide a parameter space of WD$+$He star systems for the production of double WDs that can merge and explode as SN~2021yfj-like events, and then give their binary population synthesis properties.
This paper will be organized as follows.
In Sect.\,2, we present the numerical methods and corresponding results of detailed binary evolutions, 
and the binary population synthesis (BPS) methods and results are provided in Sect.\,3. 
Finally, a discussion and summary will be given in Sect.\,4.

\section{Detailed binary evolution} \label{sec:Binary evolution methods}
\subsection{Numerical methods}
The assumptions in the stellar evolution code are similar to our previous works (see \citealp{Liu2018DD,Liu2020}).
The WDs are treated as a point mass, and the initial He star models are assumed to have a He mass fraction $Y=0.98$ with the metallicity $Z=0.02$. 
The ratio of the mixing length to the local pressure scale height is set to be 2.0, and the convective overshooting parameter is supposed to be 0.12 (e.g. \citealp{Pols1998MNRAS.298..525P}). 

In order to produce SN~2021yfj-like events via the double WD merger channel, the following conditions must be satisfied. 
First, the primary WD (WD1) must evolve to be a Si/S-rich WD. 
This requires (1) the He accretion rate should be large enough, i.e. \(\dot{M} > \dot{M}_{\rm sur\_C\_ign} =2.1 \times 10^{-6} \, M_{\odot} \, \text{yr}^{-1}\), where $\dot{M}_{\rm sur\_C\_ign}$ is the critical mass-transfer rate for the surface C ignition; and (2) the final mass of WD1 satisfies \(M_{\text{WD1}}^{\mathrm{f}} \ge 1.25 \, M_{\odot}\). 
As shown by \citet{Nomoto1985ApJ...297..531N,Saio1985A&A...150L..21S,Saio1998ApJ...500..388S,Brooks2016ApJ...821...28B,Schwab2016MNRAS.463.3461S,wang2017,wu2019,wu2020}, such accretion can trigger a surface carbon ignition that propagates inward, producing a Si/S-rich WD.
While the precise conditions for the formation of a Si/S-rich WD warrant further investigation, small variations in these conditions do not significantly affect the overall parameter space explored in this study, whereas larger changes would have a substantial impact.
Second, to ensure that WD1 is tidally disrupted during the merger rather than its companion, the mass of the secondary must satisfy \(M_{\text{WD2}} > M_{\text{WD1}}\). 
Under these conditions, the subsequent merger---driven by gravitational-wave emission in a double-degenerate system \citep{iben1984,webbink1984}---would preserve the Si- and S-rich envelope in the outer ejecta, naturally explaining the prominent Si and S emission lines, and possibly residual He features (because of the previous He-accretion process) observed in SN~2021yfj.

We explore the parameter space of WD+He star systems that can lead to the formation of SN~2021yfj-like events via the double WD merger scenario. 
In our calculations, the initial WD masses ($M_{\rm WD}^{\rm i}$) are assumed to be in the range of $1.0$$-$$1.2\,M_{\odot}$, the initial He star masses ($M_{\rm He}^{\rm i}$) in the range of $2.0$$-$$2.6\,M_{\odot}$, and the initial orbital periods ($P_{\rm orb}^{\rm i}$) ranging from $\sim0.06$ to $0.50 \mathrm{d}$.

\subsection{Detailed binary evolution results} \label{sec:binary evolution Results}
\begin{figure}[ht!]
\centering
\includegraphics[angle=270, width=\columnwidth]{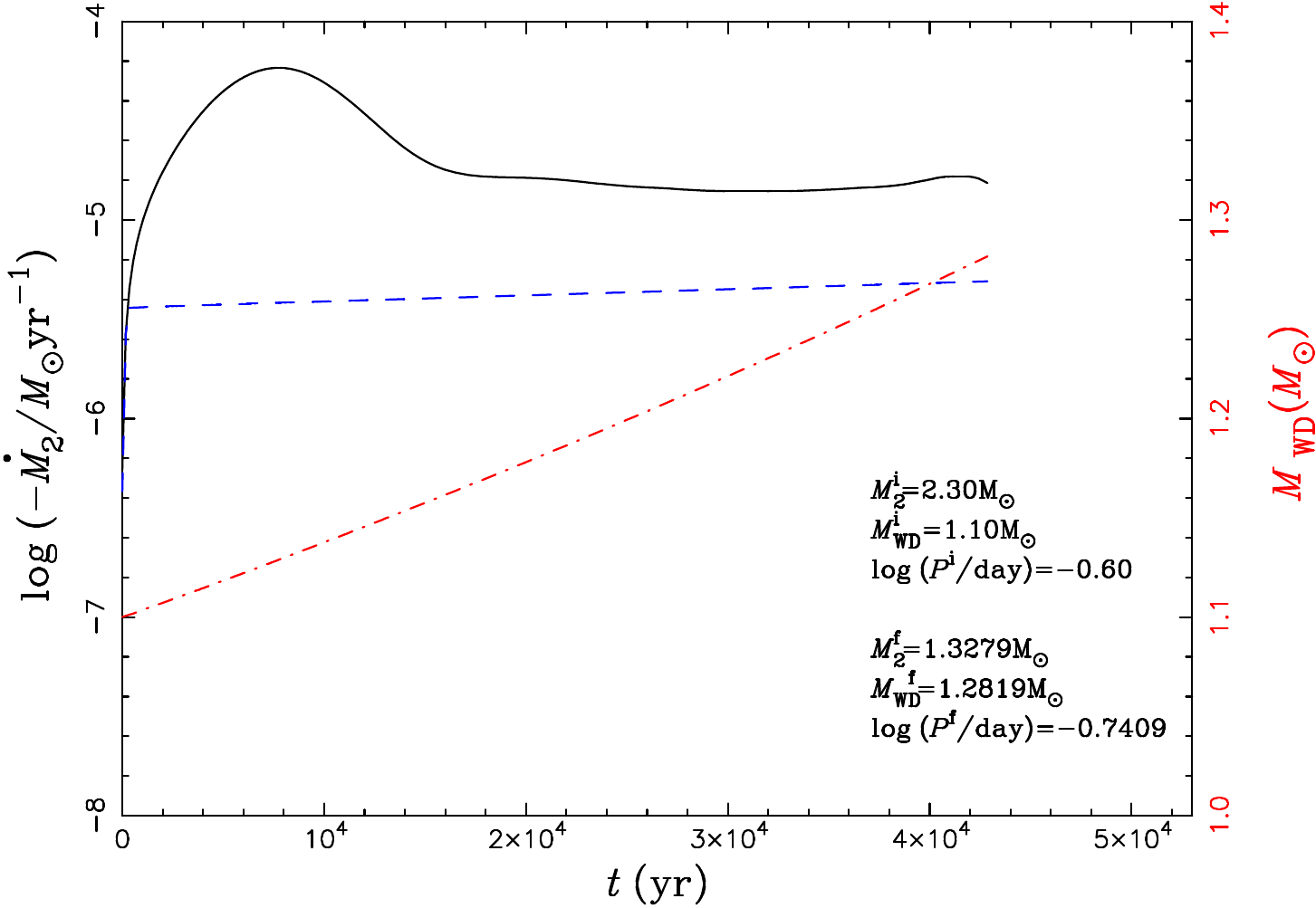}
\caption{An example for the evolution of a WD+He star system that can evolve to a double WD system with a Si/S-rich WD. 
The solid line, dashed line and dash-dotted line represent the evolution of the mass transfer rate, the WD mass-growth rate, and the primary WD mass changing with time.
\label{fig:general}}
\end{figure}

Fig.\,1 displays a representative example for the evolution of a WD+He star system that could evolve to a double WD system consisting of a Si/S-rich WD.
The initial parameters of the WD$+$He star system are ($M_{\rm WD}^{\rm i}$, $M_{\rm He}^{\rm i}$, $\log P_{\rm orb}^{\rm i}$)=($1.1\,\rm M_{\odot}$, $2.3\,\rm M_{\odot}$, $-$0.6).
From this figure, we can see that the mass-transfer rate almost always larger than $1 \times 10^{-5} \, M_{\odot} \, \text{yr}^{-1}$ after the He star fills its Roche lobe.
During this process, the mass-transfer rate is larger than the critical rate $\dot{M}_{\rm cr}$ (blue dashed line), and the transferred He-rich matter burns stably on the surface of the primary WD at the rate of $\dot{M}_{\rm cr}$, whereas the rest of the He-rich matter is assumed to be blown away from the binary in the form of optically thick wind (see \citealp{Hachisu1996ApJ...470L..97H}). 
Here, the critical rate $\dot{M}_{\rm cr}$ represents the upper limit for the stable He burning region described in \citet{Nomoto1982ApJ...253..798N}.
After this stage, the primary WD turns to be a $1.2819\,\rm M_{\odot}$ Si/S-rich WD, and the He star becomes a $1.3279\,\rm M_{\odot}$ ONe WD.
At the end of our simulation, the orbital period $\log P_{\rm orb}^{\rm f}=-0.7409$.
Subsequently, the double WDs rotate each other and will merge in about
3.3\,Gyr driven by the gravitational wave radiation, resulting in the
production of an SN~2021yfj-like event.

\begin{figure}[ht]
\centering
\includegraphics[angle=270, width=\columnwidth]{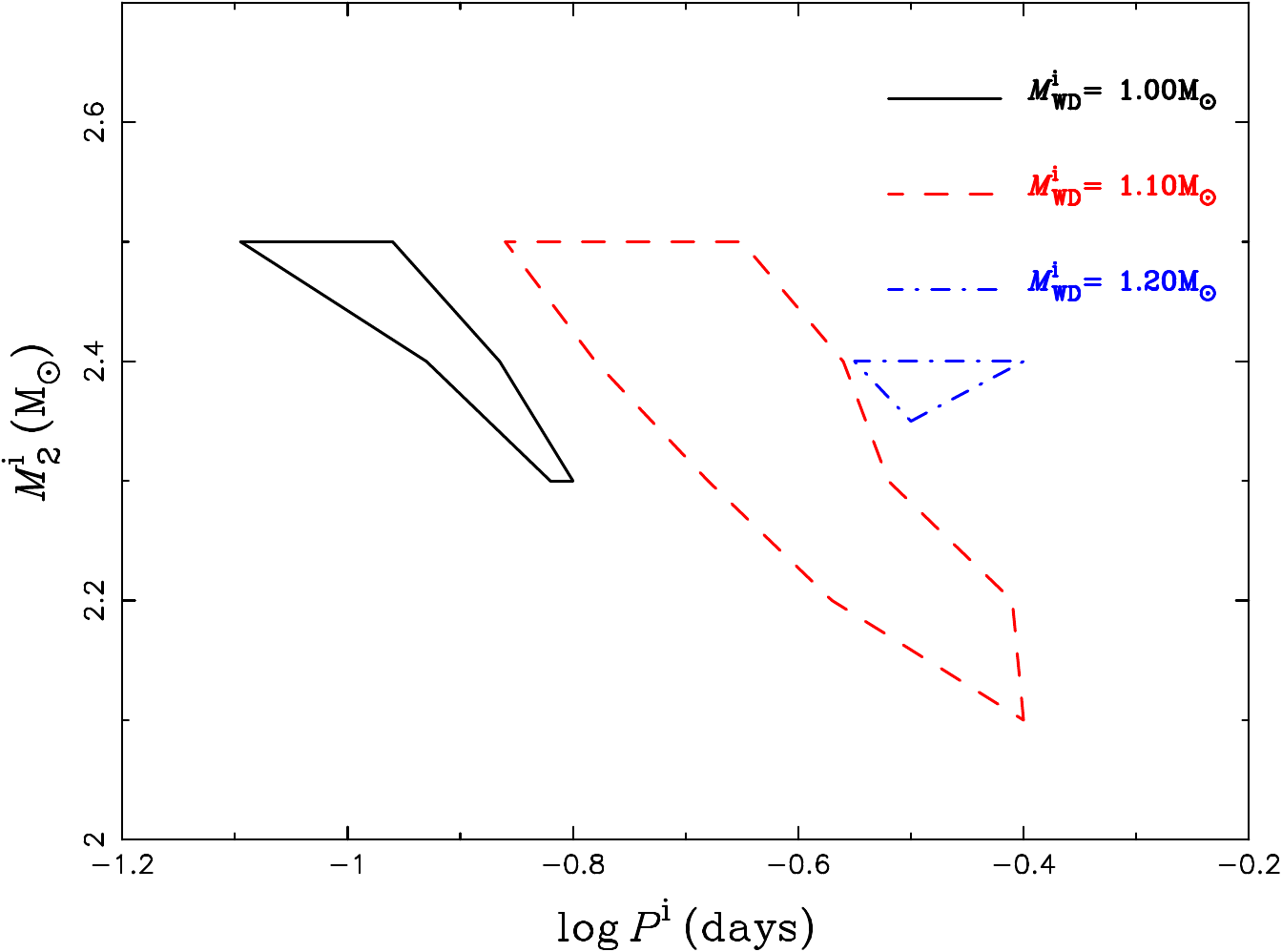}
\caption{The parameter space of WD+He star systems for the formation of SN~2021yfj-like events in the initial orbital period$-$initial He star mass ($\log P^{\rm i}$$-$$M_{\rm 2}^{\rm i}$) plane with varying initial WD masses $M_{\rm WD}^{\rm i}$.}
\label{fig:general}
\end{figure}

Fig.\,2 shows the parameter space of WD+He star systems for producing SN~2021yfj-like events based on the double WD merger scenario.
As this figure shows, in order to form double WD systems satisfying the three conditions described in Sect.\,2, the initial CO WD mass should be in the range of 1.0 to $1.2\,\rm M_{\odot}$.
The double WDs originate beyond these contours cannot produce SN~2021yfj-like events: binaries beyond the left and lower boundaries will form double WD systems in which the Si/S-rich WD is more massive than the other one, which could not evolve to mergers with Si/S-rich CSM; 
the He stars in the binaries beyond the upper boundaries are too much massive and will collapse into a neutron star; 
the WD in the binaries beyond the right boundaries cannot grow in mass more massive than $1.25\,\rm M_{\odot}$, i.e., they will only evolve to CO WD+ONe WD systems.

\section{Binary population synthesis}\label{sec:BPS methods}
\subsection{Numerical methods}

\begin{figure*}[ht!]
\centering
\begin{minipage}{0.72\textwidth}
  \includegraphics[width=\linewidth]{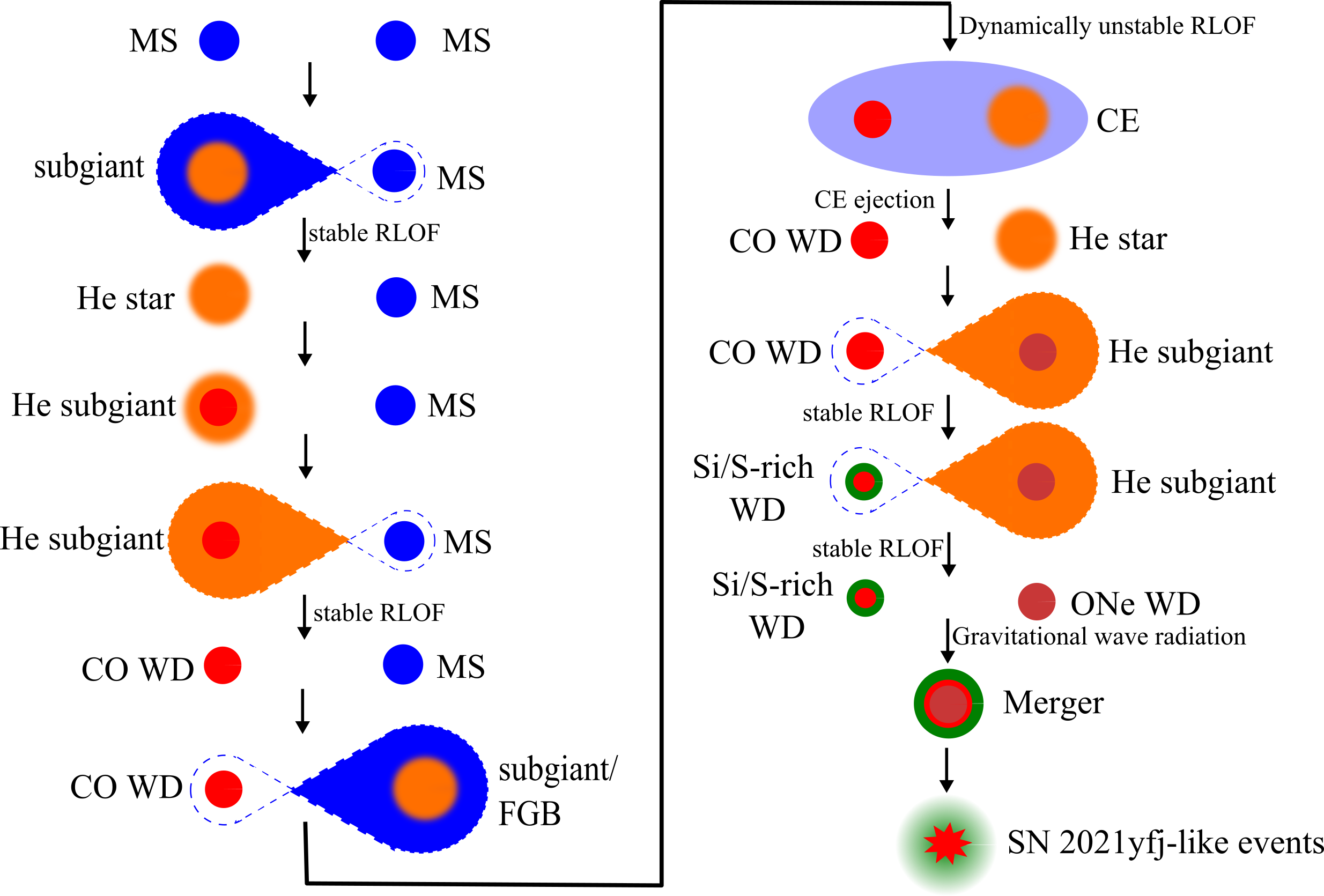}
\end{minipage}
\hfill
\begin{minipage}{0.26\textwidth}
  \caption{Evolutionary track for the formation of SN~2021yfj-like events from the double WD merger scenario.}
  \label{fig:general}
\end{minipage}
\end{figure*}

In order to investigate the statistic properties of the SN~2021yfj-like events from the double WD merger scenario, we employ the Hurley rapid binary evolution code (see \citealp{Hurley2002MNRAS.329..897H}) and perform a series of BPS Monte Carlo simulations from their formation to the formation of CO WD+He star systems. 
In each of our simulation, we evolve $10^{8}$ primordial binaries. 
An SN~2021yfj-like event is supposed to be produced if the parameters of the formed WD+He star systems are located in the initial regions obtained from the detailed binary evolution simulations shown in Fig.\,2.

The input physics and basic assumptions are similar to \citet{liu2018AIC,Liu2020}, as follows.
(1) All stars are assumed to reside in binaries with circular orbits. (2) The primary mass follows the initial mass function of \citet{Miller1979}. (3) A uniform mass ratio distribution is assumed. (4) The initial orbital separations follow the distribution described by \citet{Eggleton1989}.(5) A constant star formation rate (SFR) of $5\,M_\odot\,\mathrm{yr}^{-1}$ over the past 15\,Gyr is employed to represent spiral galaxies, while a delta-function SFR with a total mass of $10^{10}\,M_\odot$ is used for elliptical galaxies \citep[e.g.,][]{Yungelson1998, Han2004}. 
(6) The mass transfer efficiency during non-conservative RLOF phases is treated in accordance with the prescriptions described in \citet{Hurley2002MNRAS.329..897H}. 
(7) For the common envelope (CE) ejection process, which is crucial for the formation of WD+He star systems, we adopt the standard energy prescription (Webbink 1984), where the CE ejection efficiency $\alpha_{\mathrm{CE}}$ and the stellar structure parameter $\lambda$ are combined into a single free parameter. We choose $\alpha_{\mathrm{CE}}\lambda = 1.0$ for our standard model, and also consider an alternative value of $1.5$ to examine its impact.

Fig.\,3 shows the evolutionary track for the formation of SN~2021yfj-like events from the double WD merger scenario.
The primordial primary first fills its Roche lobe during its subgiant phase, initiating a stable Case B mass transfer process. 
Following this Roche-lobe overflow (RLOF), the primary will exhaust its H-rich shell and evolve into a He star. 
This He star continues its evolution and eventually expands to fill its Roche lobe again at the He subgiant stage, leading to a second stable transfer of its helium-rich envelope to the secondary.
This process results in the formation of a CO WD+main-sequence (MS) system.
Subsequently, the secondary star evolves and fills its Roche lobe at its subgiant or first giant branch (FGB) stage. 
At this point, the mass transfer becomes dynamically unstable, likely leading to the formation of a CE. 
If the CE is successfully ejected, the system emerges as a WD+He star binary. 
The future evolution of this system then proceeds along the channels detailed in Fig.\,1.
For this specific formation channel, the primordial binary parameters are in the following ranges: primary mass $M_{\rm 1,i} \sim 6.5$--$8.0 \, M_{\odot}$, mass ratio $q_{\rm i} \sim 0.55$--$0.75$, and orbital period $P^{\rm i} \sim 4$--$22$ d.
When these binaries evolve into WD+He star systems, the masses of the WDs are in the range of $1.0$--$1.2 \,\rm M_{\odot}$, the masses of the He stars range from $2.2$--$2.5 \,\rm M_{\odot}$, and the orbital periods are from 2.3 to 9 $\mathrm{h}$.

\subsection{Binary population synthesis results}\label{sec:BPS results}
In Fig.~4, we present the evolution of the Galactic birthrate of SN~2021yfj-like events as a function of time, assuming a constant star formation rate (SFR) of $5\,M_\odot\text{yr}^{-1}$. The derived Galactic rate for such events is approximately $(2.6\text{--}4.4) \times 10^{-5}\text{yr}^{-1}$, corresponding to about $1\%$ of the Galactic Type Ia supernova rate ($(3\text{--}4) \times 10^{-3}\text{yr}^{-1}$; e.g., \citealp{Cappellaro1997ASIC..486...77C, Li2011MNRAS.412.1473L}). This rate is comparable to that of other rare thermonuclear transients, such as Ca-rich gap objects (\citealp{Kasliwal2012ApJ...755..161K}), indicating that SN~2021yfj-like events are not exceptionally rare in the local universe. 
The delay-time distribution derived from a single starburst of $10^{10}M_\odot$ suggests that such events typically occur on timescales $>450$ Myr, pointing to intermediate or long delay times for this channel (see Fig.~5).
\citet{schulze2025} estimated a stellar population age of $\sim2\text{--}8\,\text{Gyr}$ for the host galaxy of SN~2021yfj, which is broadly consistent with the delay times predicted by our double WD merger scenario.

\begin{figure}[ht!]
\begin{center}
\includegraphics[angle=270, width=\columnwidth]{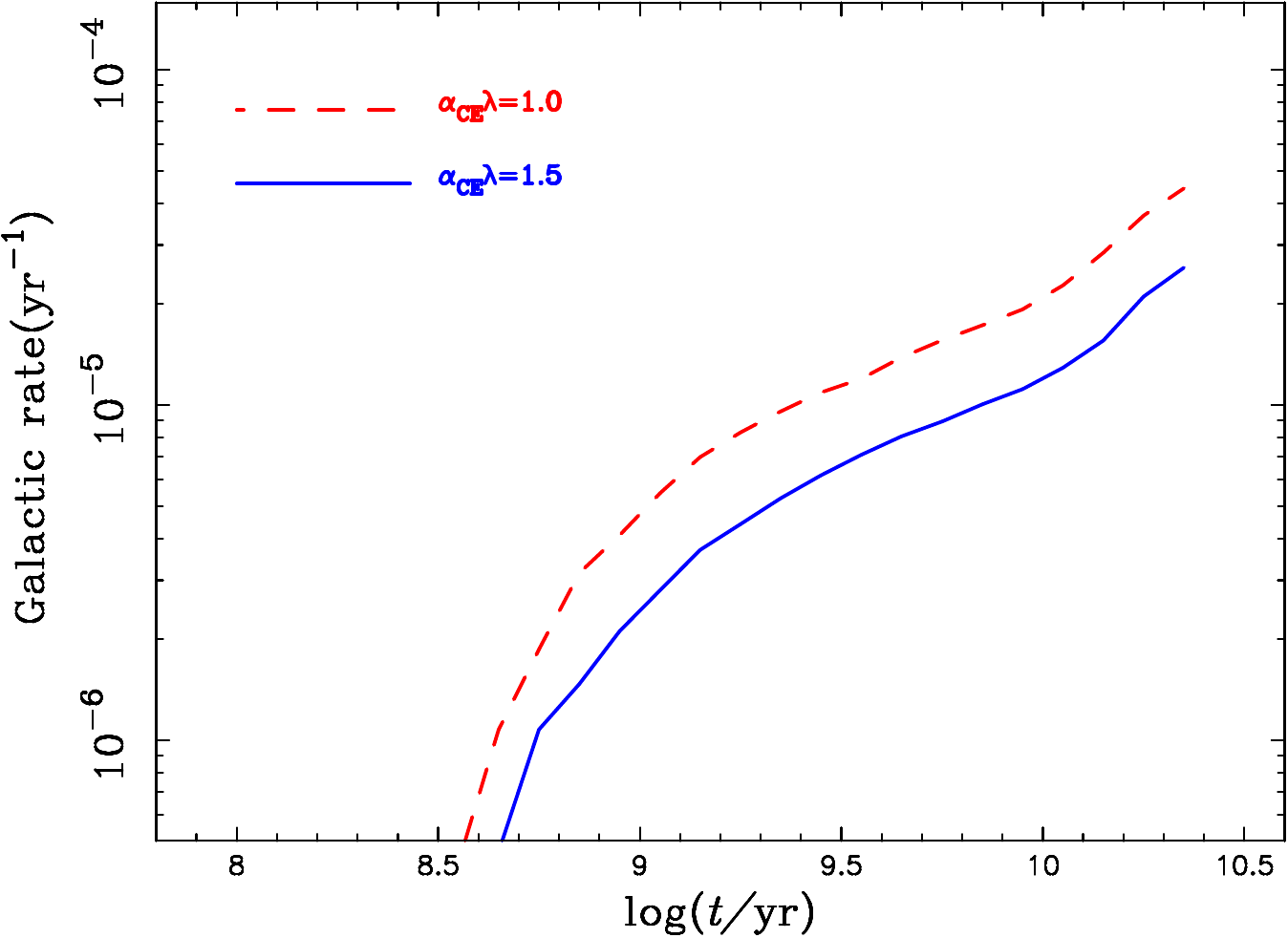}
\caption{The evolution of the Galactic rate of SN~2021yfj-like events from the double WD merger scenario. 
Here, the galactic star formation rate is assumed to be $5\,\rm M_{\odot}yr^{-1}$.
\label{fig:general}}
  \end{center}
\end{figure}

\begin{figure}[ht!]
\begin{center}
\includegraphics[angle=270, width=\columnwidth]{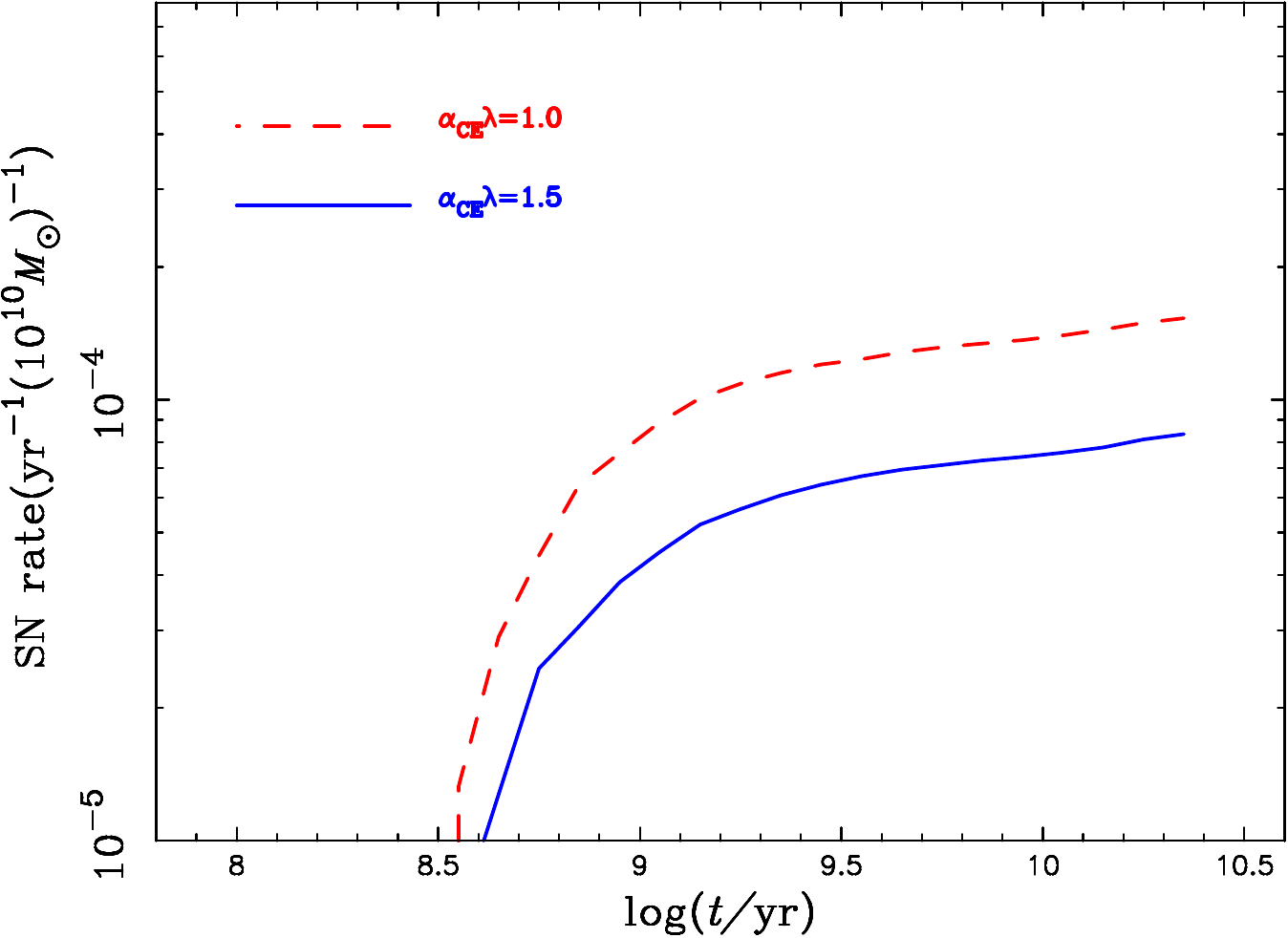}
\caption{The delay-time distributions of SN~2021yfj-like events from the double WD merger scenario.
Here, a star burst of $10^{\rm 10}\,\rm M_{\odot}$ is adopted.
\label{fig:general}}
  \end{center}
\end{figure}

Our predicted rate is higher than the observational constraints derived from the ZTF Bright Transient Survey, which has identified only one event (SN~2021yfj) over six years of operation with near-complete spectroscopic classification of transients brighter than $18.5$ mag, placing an upper limit of $<30~\text{Gpc}^{-3}\text{yr}^{-1}$ for this class (\citealp{schulze2025}). 
Several factors may account for this apparent discrepancy:
(1) The unique spectroscopic signatures with emission features of Si, S, and Ar require high-resolution spectroscopy for proper classification. 
Given that only one such event has been identified to date, it is difficult to quantify this effect precisely.
It is notable that many transient surveys do not obtain enough high-resolution spectra, implying that a non-negligible fraction of such events may have been missed or misclassified.
(2) The carbon ignition time and the mass fraction of Si, S and Ar in the Si/S-rich shell formed during off-center carbon burning vary with the He-accretion rate. This could lead to diversity in the prominence of Si, S, and Ar emission features, with events possessing lower Si/S fraction producing weaker features that may be missed or misclassified in current spectroscopic surveys.
If we conservatively assume that a Si mass fraction of $\ge 1\%$ is required to produce detectable emission features, then a significant portion of double WD mergers would yield Si/S-rich shells with insufficient Si abundance\citep{Kato2018}.
(3) The tidal stripping during the merging phase may introduce significant diversity in the mass and distribution of the CSM, potentially affecting the detectability and spectral appearance of SN~2021yfj-like events.
The mass of the stripped matter is still quite uncertain, and it seems to be the largest effect on the rate discrepancy. 
Some studies on simulating double WD mergers predict the existence of CSM during tidal stripping \citep[e.g.,][]{raskin2013,zetani2023}, while others do not \citep[e.g.,][]{pakmor2012,wu2023b}. 
It has been recently reported that SN 2020aeuh may be formed through a double WD merger with $\sim 1\, M_{\odot}$ CSM \citet{tsalapatas2025}.
Further studies on the properties of CSM during tidal stripping of double WD mergers are needed.
(4) The viewing-angle effects might also play a role; if the emission from the Si/S-rich CSM is anisotropic, only events viewed from particular orientations may display the characteristic spectral signatures. 
As shown by \citet{Suzuki2019}, emission powered by ejecta–CSM interaction in a disk-like geometry may result in a relatively large viewing angle, which means that the effect of viewing angle may be negligible.
Upcoming wide-field transient facilities, such as the Vera C. Rubin Observatory's Legacy Survey of Space and Time (LSST), have the potential to expand the sample of SN~2021yfj-like events over the next decade, which would allow for more robust statistical comparisons with our model predictions.

\section{Discussion and Summary} \label{Discussion and Summary}
We have systematically investigated the formation of SN~2021yfj-like events via double WD mergers originating from the CO WD$+$He star channel, and suggested that this is a viable and competitive model for producing such transients.
Tidal stripping of a Si/S-rich WD in this merger scenario generates CSM enriched in Si, S, and Ar during the merging process, thereby providing an alternative interpretation for the peculiar spectral features of SN~2021yfj.
Moreover, both merging WDs possess He-rich atmospheres as a result of the He-rich mass transfer that occurred prior to their formation. 
Consequently, the generated CSM would also contain a small amount of He, which can account for the weak He emission features observed in SN~2021yfj.
Compared with other possible formation models, the double WD merger scenario involves a stable mass transfer and the formation of a Si/S-rich WD, which is physically well-grounded in binary evolution theory.

A key outcome of this scenario is the formation of a WD enriched in Si, S and a small amount of Ar. 
This composition has important implications for the nucleosynthesis and spectral evolution during the merger and subsequent eruption. 
\citet{wu2020} systematically investigated the formation of such WDs through He accretion onto CO WDs, demonstrating that the surface carbon ignition can indeed produce an outer layer rich in Si and S under appropriate accretion rates ($\dot{M} > 2.1\times10^{-6}\,M_{\odot}\,\text{yr}^{-1}$) with the final primary WD mass $M_{\rm WD1}^{\rm f} \ge 1.25\,M_\odot$. 
We note that the parameter space in Fig.~2 is governed by physical boundaries related to these conditions, which carry inherent theoretical uncertainties. 
The requirement of a high He-accretion rate is easily satisfied in our models, as the relatively massive He companions ($M_{\rm He}^{\rm i} \ge 2.0\,M_\odot$) universally lead to high mass-transfer rates ($> 10^{-5}\,M_\odot\,\mathrm{yr}^{-1}$, e.g. see Fig.~1). Conversely, the threshold of the final primary WD mass ($M_{\rm WD1}^{\rm f} \ge 1.25\,M_\odot$) is the primary source of boundary uncertainty. If this mass threshold varies, it will directly shift the right-hand boundary of our parameter space: a higher (or lower) critical mass would shift the boundary to the left (or right), which would lead to a slightly lower (or higher) birthrate in our BPS model.

However, the precise structure of these hybrid WDs remains sensitive to the accretion history and the treatment of flame propagation.  
Such sensitivity has been highlighted by \citet{Kato2018}, who showed that for near-Chandrasekhar-mass WDs accreting He-rich matter, the Si yield varies significantly with the accretion rate, ranging from $<0.1\%$ to $\sim 23\%$. 
We also discussed the influence of mass-accretion process on the structure of the formed Si/S-rich WDs in our another work \citep{Wu2026ApJ..1006..156W}.
To fully constrain these outcomes, further studies are needed on multi-dimensional hydrodynamic simulations incorporating detailed nuclear networks, as well as systematic parameter studies covering different WD masses, accretion histories, and merger configurations. 
Observationally, verifying such Si/S-rich WDs is challenging because their internal composition is not directly accessible; asteroseismology of pulsating WDs could provide indirect clues, although this is rarely applied to such massive WDs.

Regarding the more massive mergers involving an ONe WD, the outcome remains theoretically uncertain: such events could lead either to a Type Ia-like explosion (e.g., \citealp{Miyaji1980,Marquardt2015,Jones2016}) or to accretion-induced collapse (AIC) into a neutron star (e.g., \citealp{Canal1990}). 
In either case, the observational signature would differ markedly from standard SNe Ia, potentially contributing to the diversity of fast-evolving and/or subluminous transients. 
Continued monitoring and spectral follow-up of future SN~2021yfj-like events will be helpful to constrain these possibilities and clarify whether the mergers of ONe WD$+$Si/S-rich WD systems contribute to this SN class.

In summary, the double WD merger scenario with a Si/S-rich WD from the WD+He star channel offers a physically coherent and observationally plausible model for SN~2021yfj-like events (or named SN Ien).
We suggest that this scenario can naturally explain the strong narrow Si, S and Ar emission features and weak He emission features in their spectrum.
The merger rate of this scenario is $\sim(2.6\text{-}4.4) \times 10^{-5}\,\text{yr}^{-1}$, which is comparable to the inferred rates of other rare thermonuclear transients, such as the Ca-rich gap objects.
We also highlight that whether an event appears as an SN~2021yfj-like transient may depend on several uncertainties, such as the timing of carbon ignition, the mass fraction of Si, S, and Ar in the WD, the efficiency of tidal stripping during the merger, and the inclination-dependent viewing angles.
To better characterize both the observational class of the SN 2021yfj-like events, further observational efforts including expanding the sample of this type transients and more detailed follow-up of SN~2021yfj-like events are needed. 
Further theoretical work remains necessary, including radiative transfer simulations to predict the observational signatures of this scenario and detailed modeling of the formation of Si/S-rich WDs and their merger dynamics.

\begin{acknowledgements}
We acknowledge useful comments and suggestions from the anonymous referee.
This study is supported by the CAS Project for Young Scientists in Basic Research (YSBR-148), the National Natural Science Foundation of China (Nos 12225304, 12288102, 12473032, 12273105 and 12090040/2021YFA1600401/12090043), the National Key R\&D Program of China (Nos. 2021YFA1600403 and 2021YFA1600400), the Yunnan Revitalization Talent Support Program (Young Talent project), International Center of Supernovae (ICESUN), Yunnan Key Laboratory of Supernova Research (No. 202505AV340004), and the Yunnan Fundamental Research Projects (Nos 202401AV070006, 202501AW070001, 202201BC070003, 202501AS070005, 202605AS350010, 202601BC070011 and 202401BC070007), the Yunnan Revitalization Talent Support Program "YunLing Scholar" project, the Strategic Priority Research Program of the Chinese Academy of Sciences (grant Nos. XDB1160303, XDB1160300, XDB1160000) and the Yunnan Revitalization Talent Support Program--Science \& Technology Champion Project (No.202305AB350003), the Grants-in-Aid for Scientific Research of the Japan Society for the Promotion of Science (JP24K00682, JP24H01824, JP21H04997, JP24H00002, JP24H00027, JP24K00668) and by the Australian Research Council (ARC) through the ARC's Discovery Projects funding scheme (project DP240101786). 
\end{acknowledgements}

\small
\bibliographystyle{aa}
\bibliography{sample701}

\end{document}